\documentclass[]{spie}  
\usepackage[]{graphicx}

\title{Few-mode fibers and AO-assisted high resolution spectroscopy: coupling efficiency and modal noise mitigation}

\author{N. Blind\supit{a}, U. Conod\supit{a}, F. Wildi\supit{a}
\skiplinehalf
\supit{a} Geneva Observatory, University of Geneva, 51, ch. des Maillettes, CH-1290 Versoix, Switzerland.
}

\authorinfo{Send correspondence to nicolas.blind@unige.ch}

\begin{document} 
  \maketitle 

\begin{abstract}
NIRPS (Near Infra-Red Planet Searcher) is an AO-assisted and fiber-fed spectrograph for high precision radial velocity measurements that will operate in the YJH-bands. 
While using an AO system in such instrument is generally considered to feed a single-mode fiber, NIRPS is following a different path by using a small multi-mode fiber (more specifically called "few-mode fiber"). This choice offers an excellent trade-off by allowing to design a compact cryogenic spectrograph, while maintaining a high coupling efficiency under bad seeing conditions and for faint stars. The main drawback resides in a much more important modal-noise, a problem that has to be tackled for allowing 1m/s precision radial velocity measurements.

We study the impact of using an AO system to couple light into few-mode fibers. We focus on two aspects: the coupling efficiency into few-mode fibers and the question of modal noise and scrambling.
We show first that NIRPS can reach coupling $\ge$ 50\% up to magnitude I=12, and offer a gain of 1-2 magnitudes over a single-mode solution. We finally show that the best strategy to mitigate modal noise with the AO system is among the simplest: a continuous tip-tilt scanning of the fiber core.
\end{abstract}



\section{INTRODUCTION}

Since its 1st light in 2002, HARPS has been setting the standard in the exo-planet detection by radial velocity (RV) measurements. Based on this experience, our consortium is developing a high accuracy near-infrared RV spectrograph covering YJH bands to detect and characterize low-mass planets in the habitable zone of M dwarfs. It will allow RV measurements at the 1-m/s level and will look for habitable planets around M- type stars by following up the candidates found by the upcoming space missions TESS, CHEOPS and later PLATO. Although several similar instruments are already operated (CARMENES\cite{quirrenbach_2016a}, GIANO\cite{oliva_2006a}) and about to be (SPIROU\cite{artigau_2014a}), NIRPS is the only one of its kind in the Southern hemisphere. NIRPS and HARPS, working simultaneously on the ESO 3.6m are bound to become a single powerful high-resolution, high-fidelity spectrograph covering from 0.4 to 1.8 micron. NIRPS will complement HARPS in validating earth-like planets found around G and K-type stars whose signal is at the same order of magnitude than the stellar noise.

Because of its AO system and that it works at longer wavelengths, NIRPS fibers works in a so-called few-mode regime, that is in-between the usual multi-mode regime (where 1000’s of modes propagates) and the single-mode. Although few-mode fibers are prone to higher modal noise, results from a significant R\&D effort made to characterize and circumvent the modal noise show that this contribution to the performance budget shall not preclude the RV performance to be achieved. 

This paper focuses on the question of coupling light into few-mode fibers with an AO-system and how to use this system to scramble modes and minimize modal noise.
In Sect.~\ref{sec:NIRPS} we give an overview of the instrument, and in particular of its AO system. NIRPS is more precisely described in [Wildi et al., 2017]\cite{wildi_2017a}.
In Sect.~\ref{sec:coupling} we briefly present the theory of coupling into few-mode fibers and apply it to the estimation of the coupling performance of NIRPS.
In Sect.~\ref{sec:scrambling} we show the principle of using an AO system to selectively couple light into different modes of the fiber as well as our current strategy to scramble light.

\section{NIRPS in a nutshell} \label{sec:NIRPS}

NIRPS (Near Infra-Red Planet Searcher) is a precision Radial Velocity (RV) instrument that will be mounted at the 3.6m-ESO telescope at La Silla Observatory by the end of 2019. It will be operated together with HARPS and will cover the near Infra-Red from 980nm to 1800nm. A later upgrade to K-band is envisioned, requiring an additional spectrograph. NIRPS is also the only planned ultra-stable high-resolution spectrograph to be installed in the Southern hemisphere.

\subsection{The science case}
NIRPS science case focuses on finding and confirming earth-mass planets in the habitable zone of low-mass M stars, in particular those identified by future space missions like TESS and PLATO. Such small planets require a radial velocity follow-up at a precision better than 1m/s. The simultaneous observation with HARPS, covering the 400-1800nm domain, will in addition help disentangling the stellar activity signal from planetary RV signal. This will be particularly important for early-to-mid-M dwarfs, where NIRPS and HARPS are expected to have a similar RV accuracy. NIRPS will also be used for the atmosphere characterisation of known transiting exoplanets, through very high-resolution transit spectroscopy.

\subsection{The instrument}

NIRPS is made of three parts: the Front End, the Fiber Link and the Spectrograph.

\subsubsection{Front End}
The Front-End function is to split light between HARPS and NIRPS thanks to a dichroic mirror, and then to inject light into NIRPS fibers. The current HARPS system will remain unchanged and will keep the exact same capabilities as today. It will be attached to the NIRPS Front-End bottom.

At the dichroic level, $\lambda$=380-690nm is transmitted towards HARPS, while $\lambda$=700-2400nm is reflected towards NIRPS. note that the dichroic can be removed to not affect HARPS-POL operations.

The Front End contains the ADC, the AO system, the fiber injection and the guiding camera of NIRPS. It is constituted of 2 arms currently:
\begin{itemize}
\item The {\bf WFS arm} uses photons in $\lambda$=700-950nm for wavefront sensing.
\item The {\bf YJH arm} uses photons in $\lambda$=980-1800nm to feed the Fiber Link and the Spectrograph. Similarly to instruments like HARPS, the light not injected into the fibers is reflected towards the NIR guiding camera.
\end{itemize}
The ADC is the only component optimized for the full range 700-2400nm for now, the rest of the optics being optimized for 700-1800nm, in particular the dichroic. These lasts will be changed to cover K-band in a future upgrade.

{\bf The AO system} operates on an on-axis natural guide star. The wavefront sensor is an OCAM2 camera in front of which is mounted a 14x14 lenslet array. The choice of the deep-depleted EMCCD OCAM2 was primarily motivated by it very good QE $\sim$ 80\% in the wavefront sensing band (700-950nm), compared to standard EMCCDs. The deformable mirror is an ALPAO DM241 in its fast version (settling time $\sim$ 0.5ms) on which 15x15 actuators are used. The interfacing and control is performed by the ALPAO Core Engine (ACE) FAST system, allowing loop frequencies up to 1kHz with as little as 0.2ms computation time delay (total delay $\le$ 1.5 frame).

While it may appear as an over-specified AO system for coupling to a multi-mode fiber, a high density of actuators is actually necessary to correct high order aberrations and inject speckles far from the fiber core back into it. With this design, we achieve an average fiber coupling of 50\% over the YJH-band up to I=12.
Although we can afford running the AO at frequencies of 250-500 Hz, running faster is enable by our hardware choices, increasing coupling by a few percent on brightest stars, and helping against potential telescope vibrations. The complete trade-off and design study of the AO system can be found in [Conod et al., 2016]\cite{conod_2016a}.

\subsubsection{Fiber Link}
The Fiber Link transports light from the Front End to the spectrograph. It is composed of two sets of fibers (each constituted of a science fiber and a reference/calibration fiber):
\begin{itemize}
\item The {\bf High Accuracy Fiber (HAF)} has an octagonal core of diameter only 29 $\mu m$, equivalent to 0.4" on sky. This fiber will be used with the AO system on the brightest targets (I $\le 12$) to reach highest RV precisions. This fiber allows to use the spectrograph at its nominal spectral resolution of 100000.
\item The {\bf High Efficiency Fiber (HEF)} has an octagonal core of diameter 66 $\mu m$, equivalent to 0.9" on sky, and will be used with faint stars. To preserve as much as possible the spectral resolution of the spectrograph, this fiber is reimaged in the double scrambler onto a 33x132$\mu m$ rectangular fiber. The spectrograph resolution is then degraded to $\sim$ 80000.
\end{itemize}
The Fiber Link also has the common function of scrambling the light and stabilizing the spectrograph LSF to allow measuring 1m/s RV signals. 
The Fiber Link ends into the cryostat, where the 4 fibers are hold in a common ferrule placed at the spectrograph entrance "slit".

\subsubsection{Spectrograph}
\begin{figure}
\begin{center}
\includegraphics[width=0.5\textwidth]{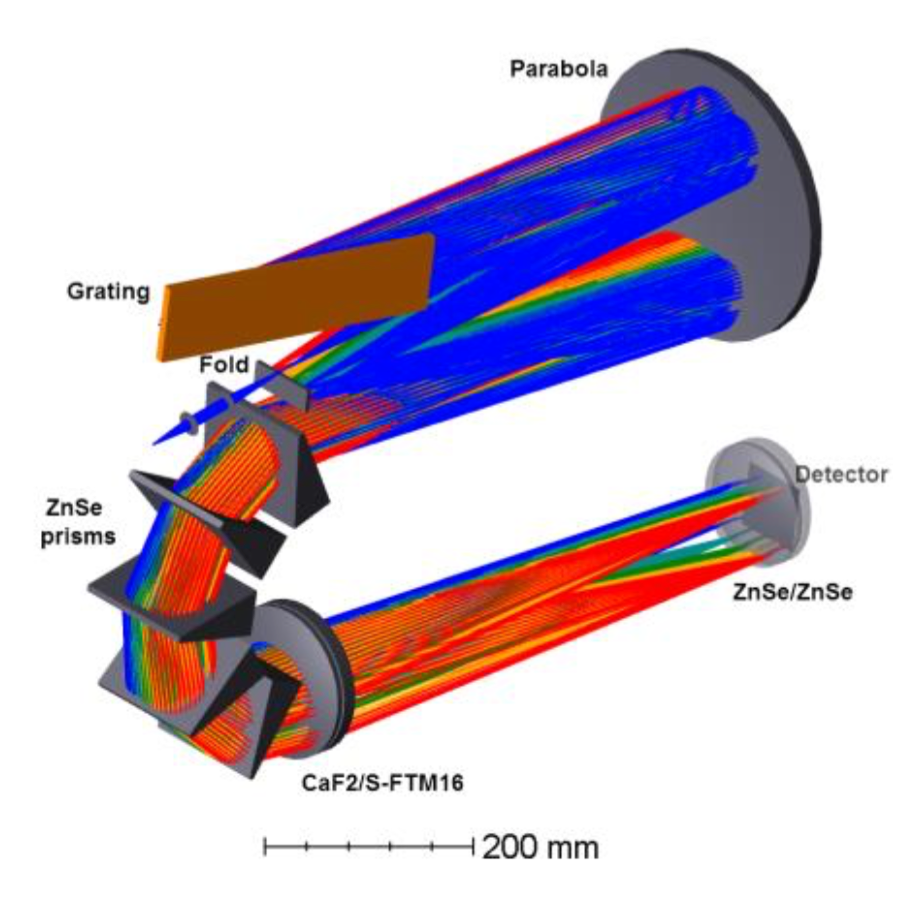}
\caption{View of the spectrograph optica design. The  length is about 1200mm, twice less than HARPS for instance.} \label{fig:spectrograph}
\end{center}
\end{figure}

The spectrograph has a white pupil design (Fig.~\ref{fig:spectrograph}): light from the fibers is first collimated on the R4 grating where it is dispersed at high spectral resolution. A series of 5 ZnSe prisms is then used as cross disperser. The input fibers are finally reimaged onto an H4RG detector, hence covering the spectral range from 980 to 1800nm. The spectrograph nominal spectral resolution is 100000 when used with the HAF. With the HEF, the spectral resolution degrades to about 80000.

Thanks to the AO system, enabling to reduce the fiber size by a factor of 2, the spectrograph is very compact: it only measures 1200 mm on its longest side, and is twice smaller (linearly) than HARPS. The cryostat itslef is a cylinder measuring 3400mm x 1120mm. This presents an important advantage to stabilize it to 1mK, while significantly reducing cooling cycles of the cryostat during AIT.

%

\section{COUPLING EFFICIENCY IN FEW-MODE FIBERS}  \label{sec:coupling}

\subsection{Coupling theory and coupling performance}

\begin{figure}
\begin{center}
\includegraphics[width=0.6\textwidth]{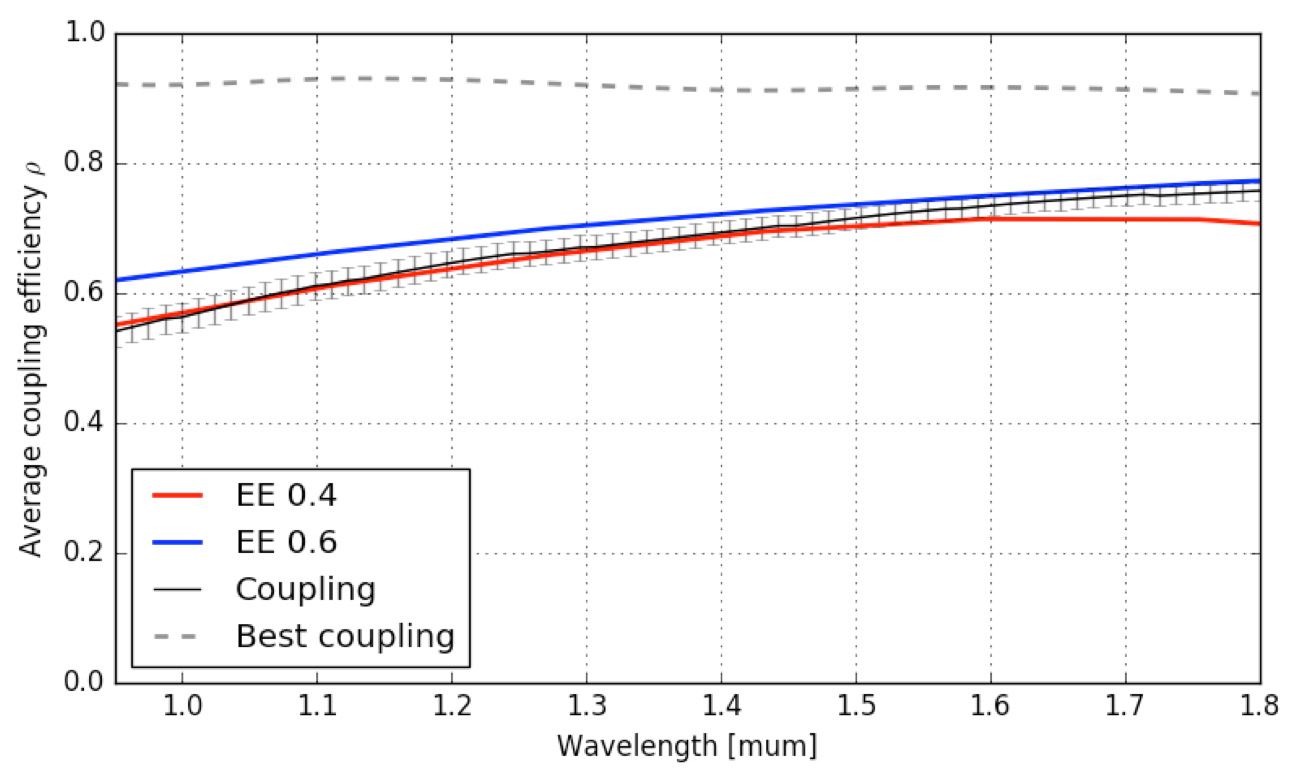}
\caption{NIRPS expected fiber coupling (black) compared to EE in 0.4” and 0.6” for a bright star and median seeing. Dotted line shows the maximum expected coupling without turbulences.} \label{fig:coupling}
\end{center}
\end{figure}

The number of modes that can transport a circular waveguide is roughly equal to $V^2/4$, where $V$ is the fiber parameter:
\begin{equation}
V = 2\:\pi\: \mathrm{NA} \:a \:/\: \lambda
\end{equation}
where $NA$ is the fiber numerical aperture, $a$ the core radius and $\lambda$ the operating wavelength. With the choice 

NIRPS High Accuracy fiber has a core size diameter of 29$\mu m$ and only transports 10 to 35 modes, hence working in the so-called few-mode regime. In this regime, geometrical optics does not apply anymore: we must compute the coupling $\rho_{lm}$ into every individual mode $lm$ of the fiber as the overlap integral between those modes and the instantaneous electric field in the telescope pupil $E_{tel}$\cite{horton_2007a}:
\begin{equation} \label{eq:coupling}
	\rho_{lm} = \frac{|\int E_{tel} E_{lm}^* dA|^2}{\int |E_{tel}|^2 dA \quad \int |E_{lm}|^2 dA}
\end{equation}
The total coupling is then simply the sum of all individual couplings:
\begin{equation}
	\rho = \sum_{l,m} \rho_{lm}
\end{equation}
We have extended the work of Horton et al.\cite{horton_2007a} (which only considers the case of a diffraction limited sources) to the case of an AO-assisted ground-based telescope. By injecting the wavefront residual phases $\phi_{res}$ obtained from CAOS simulations so that $E_{tel} \rightarrow E_{tel} \times \exp(j \phi_{res})$, we computed the coupled energy into the few-mode fibers. 

We show in Fig.~\ref{fig:coupling} the result of such computation, for median seeing (0.9") and a bright star (I $\le$ 10). We compare the coupling to the corresponding Encircled Energy (EE) and show that they are both equivalent for $N_{\mathrm{mode}} \ge$ 15 ($\lambda \le$1.5 $\mu$m for NIRPS). For $N_{\mathrm{mode}} \le $ 15 ($\lambda \ge$ 1.5$\mu$m in our case), we observe that $\rho \ge$ EE: in this condition, the modes are not totally constrained into the fiber core but are guided through a small section of the fiber cladding. Such modes are therefore also able to couple speckles that fall in the cladding, within an area of diameter 0.5-0.6''.  We can also note that the coupling of NIRPS HAF is expected to exceed that of an equivalent seeing-limited instrument with a twice as large core, up to I=11-12 (depending on the seeing). The simulations show that degrading seeing from 0.7" to 0.9" and then to 1.2" is roughly equivalent in terms of coupling efficiency to loosing 1 magnitude on the guide star for a given seeing.

The HEF fiber will be a seeing-limited fiber for fainter stars that will provide spectra with a degraded resolution of $\sim$ 80000. This loss of resolution subsequently degrades the quality factor of the RV signal\cite{bouchy_2001a, figueira_2015a}, so that this fiber will be more efficient (in terms of radial velocity) than the AO-assited HAF one for I $\ge$ 12-13.

\subsection{Sensitivity to aberrations} \label{sec:aberrations}

\begin{figure}
\begin{center}
\includegraphics[width=1.\textwidth]{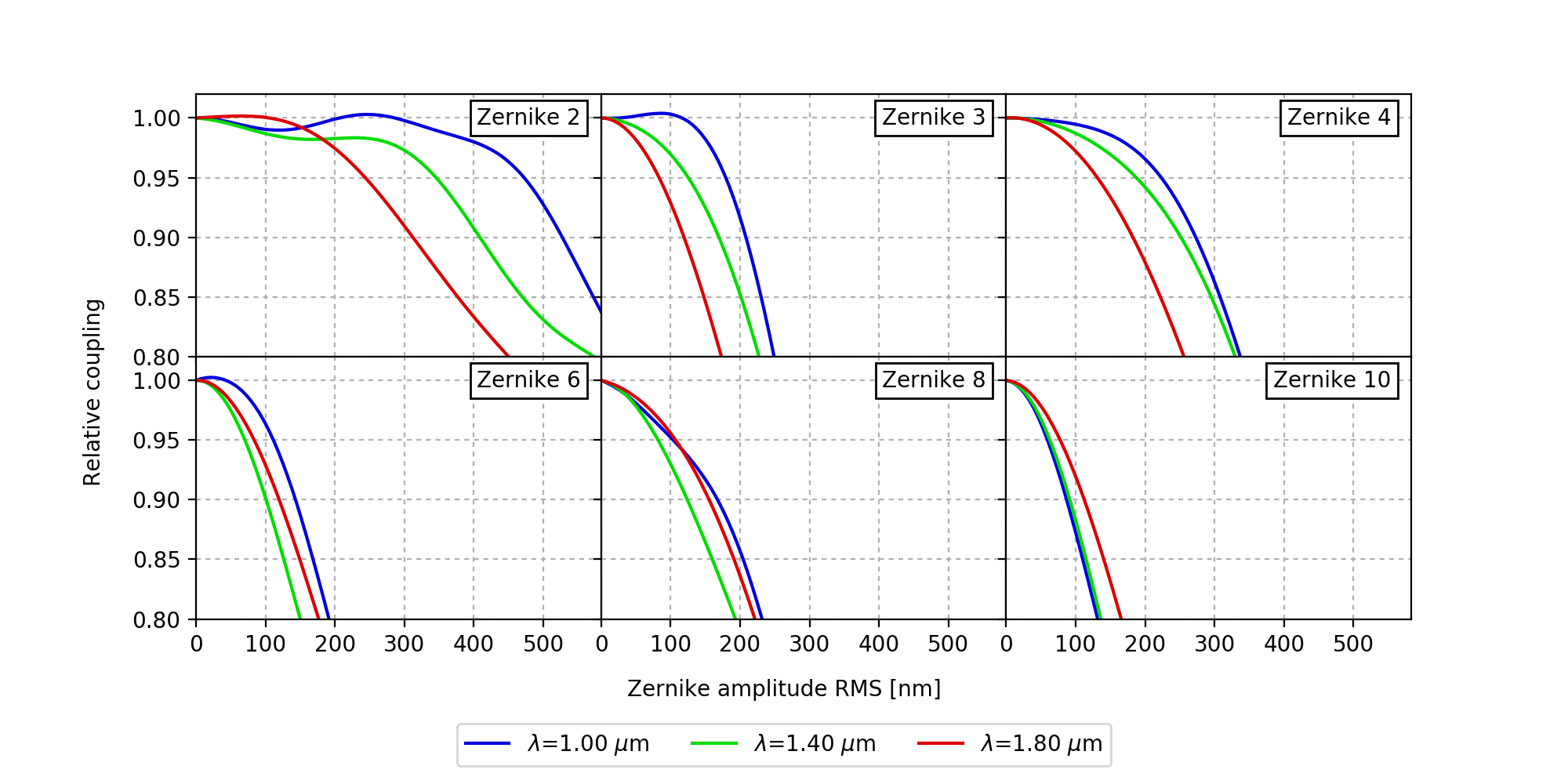}
\caption{Effect of various Zernike aberration to fiber coupling degradation in the HAF. Relative coupling of 1 correspond to the case with no NCPA.} \label{fig:coupling_vs_zernike}
\end{center}
\end{figure}

In the multi-mode case, there is no simple relation between the coupling efficiency $\rho$ and the Strehl ratio, as can be found in the single-mode fiber case \cite{coudeduforesto_2000}. While single-mode coupling is very sensitive to instantaneous wave-front error, we did not observe such a behavior in the few-mode case. A small correlation with the RMS slopes can be observed though. Each fiber mode having a particular spatial distribution (and more specifically, we shall consider the mode's phase, more than its amplitude), it will react differently to a particular aberration. Hence while a given mode will couple less,  another one can couple more. We show in Fig.~\ref{fig:coupling_vs_zernike} the impact of injecting different Zernike aberrations (corresponding to NCPA for instance). The behavior of these curve shows little dependence on the wavefront (perfectly flat or an AO residual) before introducing NCPA. The behavior with wavelength is also not intuitive: tilt and defocus are more critical at longer wavelength, which can be explained by a lower modal content with increasing wavelengths.

Coupling performance eventually appears more sensitive to terms of azimuthal order 0 like focus (\# 3) or sphericity (\# 10). Their azimuthal symmetry does not allow coupling in high order modes (at least for the perfect fibers we simulate here), conversely to e.g., coma which can increase coupling into the LP11 mode or instance. As a rule of thumb, it appears that 5\% coupling loss is achieved for NCPA amounting to 100 nm RMS.

\subsection{Figure of merit for coupling efficiency and encircled energy}

For an instrument like NIRPS, we are interested into coupling efficiency, or, quite equivalently as we have seen, to encircled energy. We can wonder whether there is an optimal "AO control criterion" that optimizes encircled energy?
Following the last two sections, it appears that EE/coupling is actually optimized for a PSF approaching the diffraction limit. Any kind of optical aberration simply reduces coupling.  So, if we are only interested in maximizing coupling, we can equivalently maximize the Strehl ratio.

As mentioned earlier, in the case of high precision radial-velocity, a second aspect to consider is modal noise, which requires a different approach. We discuss it in Sect.~\ref{sec:scrambling}.

\subsection{Comparison to the Single-Mode solution}

We can enumerate a few advantages offered by a multi-mode fiber over a single-mode one:
\begin{itemize}
\item Higher coupling efficiency, with a given AO system.
\item Higher magnitude limit, thanks to relaxed AO requirements.
\item Lower sensitivity to seeing variations and AO correction quality (i.e.~seeing and guide star magnitude);
\item Lower sensitivity to NCPA: while 100nm RMS NCPA would lead to 10-35\% losses in a single-mode fiber in YJH band, it is limited to 5\% in a few-mode fiber.
\end{itemize}
The main drawback of the choice made for NIRPS resides in the LSF stability: with only 10 to 35 modes, it will be among the astronomical instrument with the most important modal noise possible (Sect.~\ref{sec:scrambling}). A particular effort has therefore been made to study modal noise in NIRPS and to develop new scrambling strategies and demonstrates that it will indeed be able to reach a 1m/s RV long term precision [Blind et al., in prep.]. It will rely on several devices (e.g.~double scramblers), in which we expect to lose about 30\% of the light, hence lowering the total throughput advantage of NIRPS over an equivalent single-mode solution.  

Considering this 30\% losses for NIRPS Fiber Link, we estiate the gain of the multi-mode solution still 1 to 2 magnitude over the single-mode one thanks to the lower sensitivity of few-mode fibers to aberrations.The gain here mostly depends on whether we apodize the telescope pupil \cite{jovanovic_2017b} or not respectively. If we don't, the central obscuration of the 3.6m telescope ($\sim 30$\%) limits single-mode coupling efficiency to $\rho_0$ = 60\%.

Another advantage of using a single-mode fiber is that the spectrograph can be extremely compact (see for instance \cite{feger_2016a}). However, in the eventuality that a seeing-limited fiber is required for fainter objects, the compactness cannot be maintained because of the increase in \'etendue.

\section{SCRAMBLING WITH AN AO SYSTEM}
\label{sec:scrambling}

\subsection{Modal noise and scrambling}

Modal noise is the instability in the fiber output illumination leading to instability of the spectrograph Line Spread Function (LSF), and eventually radial velocity noise. Any disturbances at the fiber entrance (e.g.~guiding errors, vibrations) or along it (e.g.~varying bend stress during tracking) can be transported up to the fiber output, hence generating a change of the spectrograph LSF. Losing this information to get a perfectly stabilized LSF is called scrambling.

Modal noise originates from the combination of two effects:
\begin{enumerate}
\item An incomplete filling of the fiber near- and/or far-field;
\item Intermodal phase variations, corresponding in geometric optics to the difference of optical path traveled by rays with different arrival angles.
\end{enumerate}
The combination of those effects generate at the fiber output a speckle pattern (the interference between modes) that contains to some degree information on the injection conditions and local environment. Therefore this speckle pattern varies with time. Those effects in addition slowly evolve with wavelength, requiring bandpass of several 10nm to average out to a level of 1 m/s.

In the case of NIRPS, modal noise is increased by two factors. First, the number of modes guided by the fiber intrinsically smaller because of longer wavelength and smaller core. Second, achieving high coupling requires high Strehl ratio, which only excite a limited number of modes. In case the PSF is centered on the fiber core for instance, the modes of azimuthal order 0 are preferentially excited, and only 30 to 40\% of the modal content is excited. 

Since the Fiber Link does not have a fixed shape and is not thermally stabilized, it is necessary to include in our AO control strategy a way to deal with effect \#1.

\subsection{Mode-selective coupling} 

\begin{figure}
\begin{center}
\includegraphics[height=7.3cm]{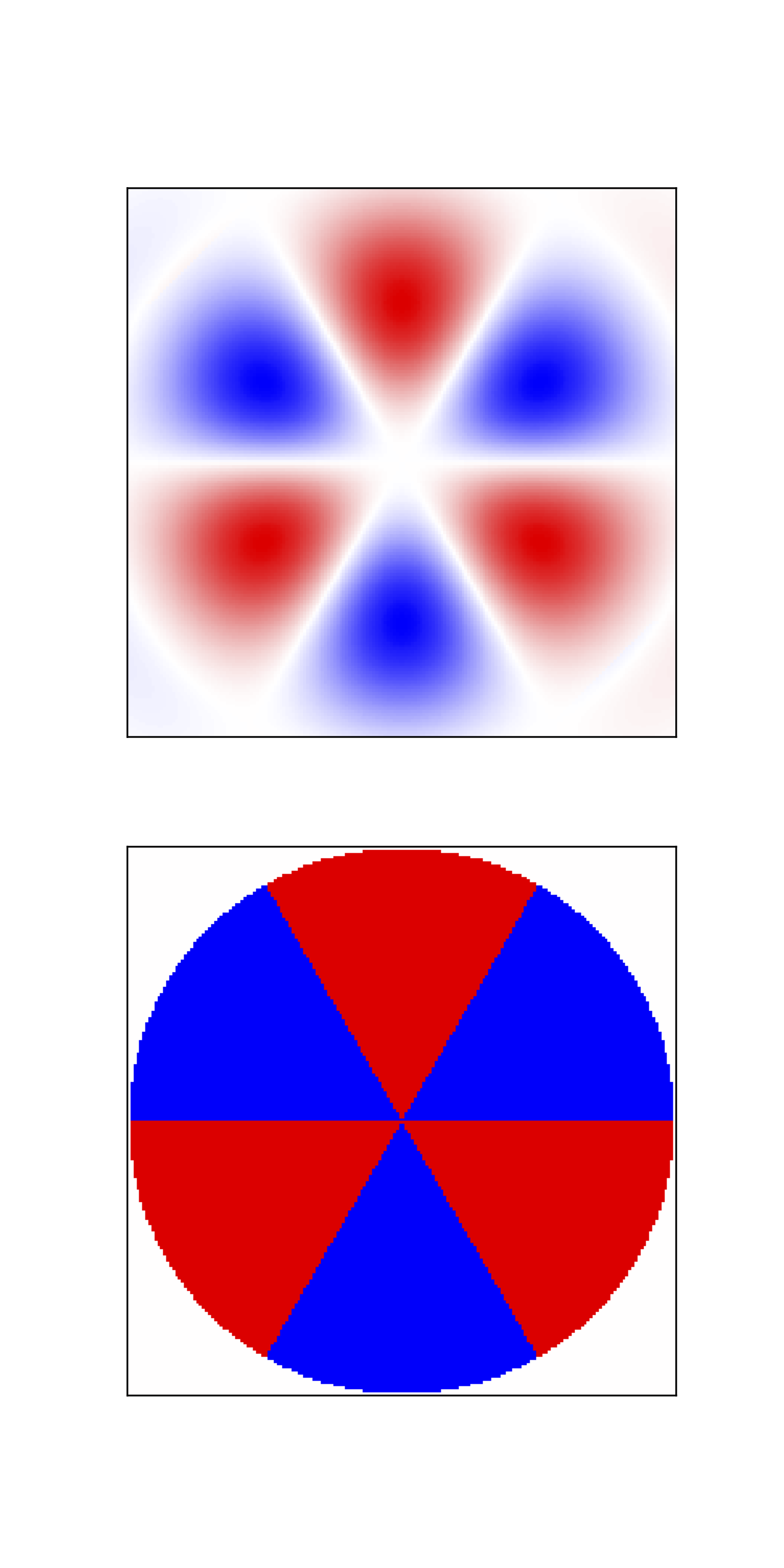} \hfill
\includegraphics[height=6.5cm, trim={1cm 0 2cm 1.4cm}, clip]{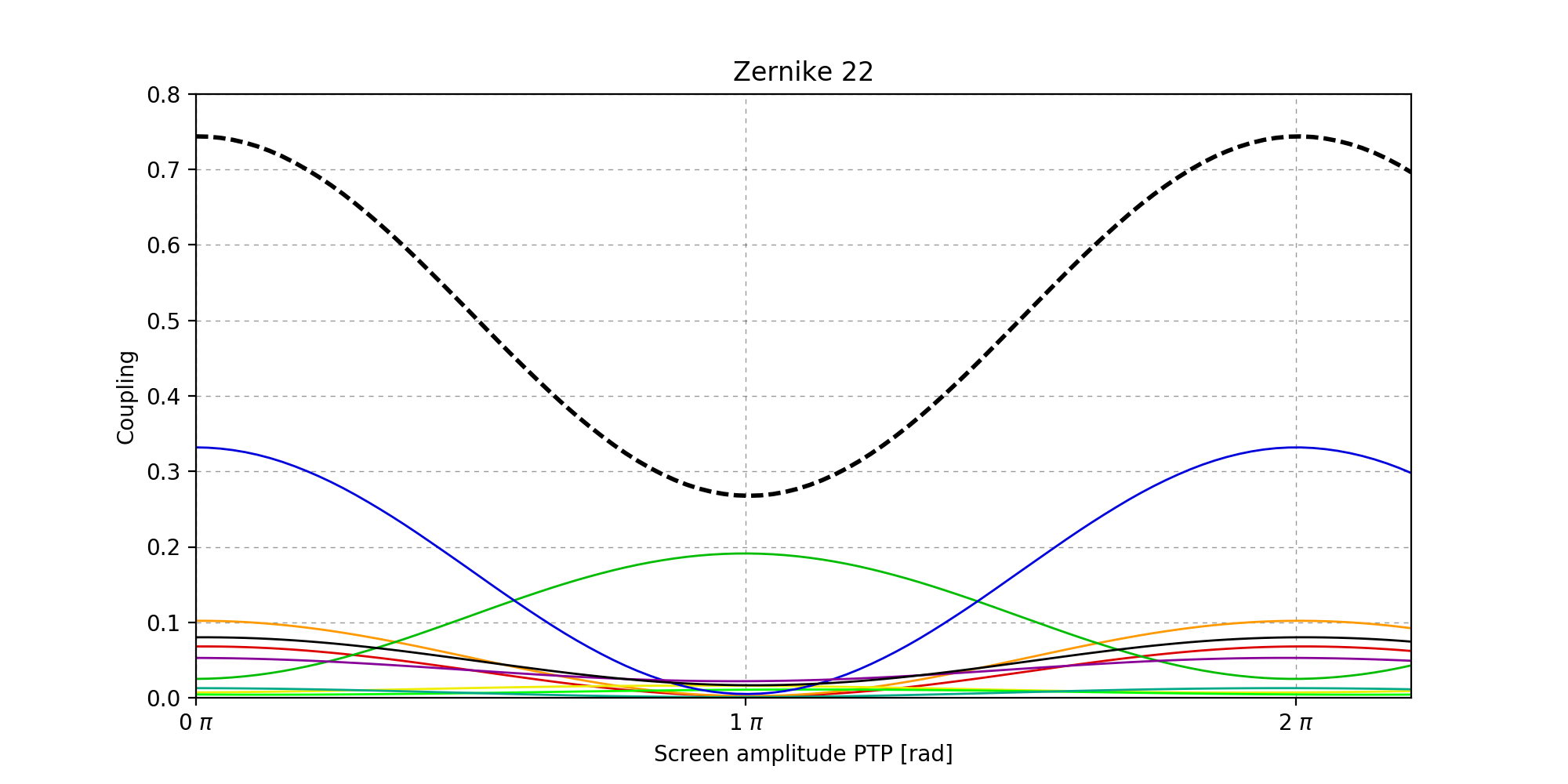}
\caption{Mode selective coupling into LP31. Top left: LP31 mode electric field in the far-field. Bottom left: corresponding phase of LP31, applied as a phase screen onto the DM. Right: resulting coupling. Each solid line represents coupling into a different mode, the green one corresponding to LP31. The black dotted line corresponds to the total coupling.} \label{fig:selective_coupling}
\end{center}
\end{figure}

Following equation \ref{eq:coupling}, it is clear that to perfectly couple light to a mode (and only one), we must be able to match both the amplitude and phase of the telescope electromagnetic field to the one of the given mode. If we limit ourselves to the case of a standard AO system, we can only match phase. Fig.~\ref{fig:selective_coupling} shows an example of mode-selective coupling for the mode LP31, where it works particularly well. To optimize coupling to this mode, we need to apply with the DM a phase corresponding to the one of the mode in the far-field. The latter being 0 or $\pi$, we optimize (minimize) coupling to the given mode for a phase amplitude of $\pi$ (0 respectively). This is shown in the right plot for LP31 (green curve). We however see that the price to pay is very high, with total coupling degrading by a factor of more than 2 (dotted, black curve): optimizing coupling on a single mode puts out of phase many other modes.

The example shown here is rather favorable for mode selection: other modes can be more difficult to couple this way and will decouple much more other modes, i.e. will decrease total coupling even more. In practice, such a strategy may be difficult to apply. First, because of local stresses at fiber entrance, modes are generally asymmetrical and should be properly characterized, ideally one-by-one. In addition, DM actuators are (generally) positioned on a square matrix, which can not allow to apply such sharp screens for all modes. Finally, this strategy can hardly apply to broadband instruments, the number of modes and their geometry evolving with wavelength.

\subsection{Scrambling control strategy}

Because the cost to pay for mode-selective coupling is too high, we investigate a more conventional solution by injecting Zernike aberrations instead. 

Like we did in previous sections, we use a sequence of AO corrected wavefronts and add on top different Zernike aberrations with various amplitudes. We then consider that a mode will contribute to stabilizing the LSF if its interference with other modes is bright enough compared to the brightest modes. We consider a mode as "bright" when its coupling exceeds 10\% of the brightest mode. This corresponds to a "speckle contrast" $\sim$50\% which increases the weight of a particular mode to modal noise. This is a crude estimator, prone to over estimating the importance of a mode: we should ideally consider the spatial distribution of the various modes. However, the number of modes counted this way only evolves significantly for tip-tilt. In addition tip-tilt is the only modes that keeps high coupling efficiency. Defocus and higher order aberrations quickly degrade coupling by more than 10\% (as seen in Sect.~\ref{sec:aberrations}), while a very limited sample of new modes is excited. 

\begin{figure}
\begin{center}
\includegraphics[width=0.8\textwidth]{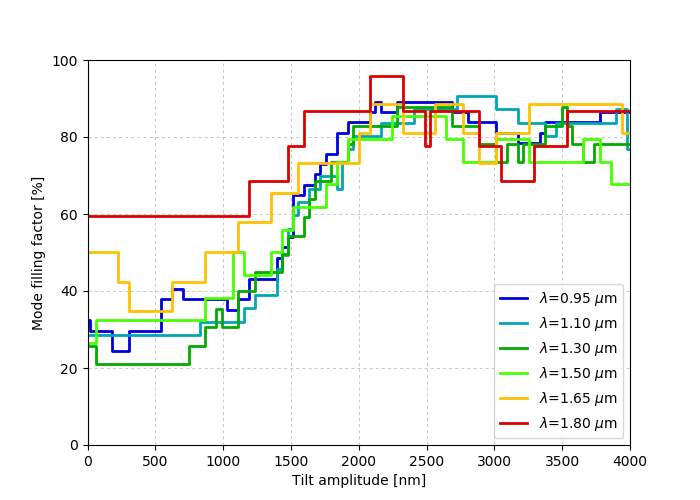}
\caption{Coupling factor for various tilt up to the edge of the NIRPS fiber core (tilt amplitude of 3.5 $\mu m$). We count all modes excited by a circular scanning of given radius.} \label{fig:coupling factor}
\end{center}
\end{figure}

Hence, tip-tilt appears as the only efficient way to scramble with the AO system. More interesting, these simulations show that no higher order aberration is able to excite a mode that tip-tilt is not able to excite with higher efficiency.  Fig.~\ref{fig:coupling factor} finally shows that up to 90\% of the modes can be excited for a given amount of tip-tilt. With a continuous scanning of the fiber head with varying distance from the core center, we can potentially fill all modes of the fiber in a controlled manner.

We could recently apply this strategy on our dedicated fiber test bench in Geneva. We used a 25 $\mu m$ core size fiber (Thorlabs FG025LJA, which is the closest commercial fiber from the high accuracy NIRPS one), whose core is scanned thanks to an XY translation stage. The test was performed with various narrowband filters ($\Delta \lambda$ = 10nm) of central wavelength between 950 and 1650 nm and we observed a consistent improvement in LSF stability by a factor of 2-3 for all of them. The final AO system components have now been delivered to Geneva and are under test and integration. We will soon be able to verify the various injection strategies that have been explored in the last two sections.

%

\section{CONCLUSION}

In this paper we present preliminary work that aims at understanding the coupling of an AO system to a few-mode fiber, in the framework of high precision radial velocity measurements. This new approach led us to reconsider our AO control strategy in order to optimize at the same time fiber coupling and scrambling (or in other words, spectrograph line stability).

We have first shown that high coupling efficiency can be achieved on faint stars ($\rho \ge$50\% over YJH up to I=12) by using few-mode fibers. In particular, the comparison to an equivalent single-mode solution shows a lower sensitivity to aberrations (atmosphere residuals and NCPA), hence allowing relaxed specifications of the AO system and a higher magnitude limit of 1-2 magnitudes in favor of the few-mode solution.

The drawback of using few-mode fibers is a very high modal noise, which appears to be the main contributor to our radial velocity error budget. We have defined an optimal AO control strategy that consists in a continuous tip-tilt scanning of the fiber core during the observations. Although we show that mode-selection is possible to some extent with a classical AO, it involves unacceptable losses.

A test bench is now assembled in Geneva with the final NIRPS AO system. In the coming months, we will verify the theoretical work presented in this paper and evaluate what is effectively the best control strategy for optimizing both coupling and scrambling.


\bibliography{/Users/nblind/Travail/Papers/biblio}   
\bibliographystyle{spiebib}   

\end{document}